# Enhancing single photon emission through quasi-bound states in the continuum of monolithic hexagonal boron nitride metasurface

Shun Cao[1], Yi Jin[1], Hongguang Dong[1,2], Tingbiao Guo[1], Jinlong He[3] and Sailing He[1,2,4*]

[1]*Centre for Optical and Electromagnetic Research, National Engineering Research Center for Optical Instruments, Zhejiang University, Hangzhou 310058, China*

[2]*Ningbo Research Institute, Zhejiang University, Ningbo 315100, China*

[3]*College of Modern Science and Technology, China Jiliang university, Hangzhou, 310018, China*

[4]*Department of Electromagnetic Engineering, School of Electrical Engineering, KTH Royal Institute of Technology, S-100 44 Stockholm, Sweden*

*sailing@kth.se

## Abstract

A patterned structure of monolithic hexagonal boron nitride (hBN) on a glass substrate, which can enhance the emission of the embedded single photon emitters (SPEs), is useful for onchip single-photon sources of high-quality. Here, we design and demonstrate a monolithic hBN metasurface with quasi-bound states in the continuum mode at emission wavelength with ultrahigh Q values to enhance fluorescence emission of SPEs in hBN. Because of ultrahigh electric field enhancement inside the proposed hBN metasurface, an ultrahigh Purcell factor (3.3 × $10^4$) is achieved. In addition, the Purcell factor can also be strongly enhanced in most part of the hBN structure, which makes the hBN metasurface suitable for e.g. monolithic quantum photonics.

## 1. Introduction

Recently, hexagonal boron nitride (hBN) has attracted much attention as an emerging material, offering novel properties for electro-optical, optical and quantum photonic applications [1-3]. Many studies have been carried out to extensively exploit its ability to host ultra-bright quantum SPEs, even at room temperature [4]. Therefore, it is important to further study the emission enhancement of these hBN SPEs via the Purcell effect. One can integrate hBN SPEs with plasmonic or photonic nanostructures to achieve such an emission enhancement, due to the brightness and stability of hBN SPEs [5-9]. By coupling hBN SPEs to plasmonic gold nanospheres, a high fluorescence enhancement of over 300% was achieved in experiments due to high electric field concentrations and small mode volumes [5]. Tran *et al* proposed to hybridize hBN with high-quality and low-loss gold nanocavity arrays to enhance the spontaneous emission rate [6]. Except for plasmonic nanostructures, tunable optical dielectric cavities, waveguides and photonic crystal cavities (PCCs) were also used to integrate with hBN SPEs to enhance the single-photon emission [7-9]. However, the above hybrid approaches are inherently limited by the spatial mismatching between the SPEs and the electric field hotspots. Hence, spatial overlap between the quantum emitters and the electric field maximum of optical modes is essential to realize optimal coupling.

The unique optical properties of hBN make it a fascinating candidate to be fabricated directly in a monolithic system. In particular, hBN maintains transparent in the visible range for its wide optical bandgap [10]. Furthermore, hBN exhibits high chemical stability and excellent thermal conductivity, which are beneficial to micro/nano fabrications [11]. Recently, hBN has already been used as a parent material to fabricate PCCs, micro-ring resonators, etc [12,13]. Although PCCs of high Q can be realized from the monolithic hBN layers, the Purcell enhancement was not as high as expected and one should precisely position the SPEs to electric field intensity hotspots of cavity modes.

In recent years, bound states in the continuum (BICs), which represent localized resonance modes embedded in radiative continuous spectra, have received many interests [14,15]. Kolodny *et al* reported that the enhancement of the Purcell factor in pillar microcavities can be realized by employing quasi-BIC modes [16]. However, this method requires top and bottom Bragg mirrors and the other material, and thus is more complex to fabricate and the SPEs should be precisely put into the location of maximal electric field intensity. High-Q Fano resonances originated from quasi-BICs can boost the light confinement inside the metasurfaces, which greatly enhance nonlinear harmonic conversion efficiency [17]. However, there are few studies to enhance internal SPEs by applying BICs or quasi-BICs. As a consequence, a hBN metasurface with BIC or quasi-BIC modes should be designed carefully for the Purcell enhancement of SPEs.

In this work, we design and demonstrate a monolithic hBN metasurface with quasi-BIC mode at emission wavelength to enhance fluorescence emission of SPEs in hBN. The Purcell factor of the SPEs in an hBN metasurface can be tuned by changing the asymmetry parameter. An ultrahigh Purcell factor ($3.3 \times 10^4$) is achieved. In addition, the Purcell factor can also be strongly enhanced in most parts of the patterned hBN structure, which can ease experimentally the precise positioning of the SPEs in the patterned hBN structure.

## 2. Methods

A TE-polarized plane wave normally incidents onto the metasurface, as shown in Fig. 1(a). The numerical analysis of the reflectance and eigenmode spectra for the designed metasurface with different asymmetry parameter δ (from 0 to 0.1) are performed based on commercial finite element technique (COMSOL Multiphysics). The hBN was modeled as a uni-anisotropic material in this study and the used birefringence refractive indices of hBN are $n_x = n_y = 1.84$ and $n_z = 1.72$ [12].

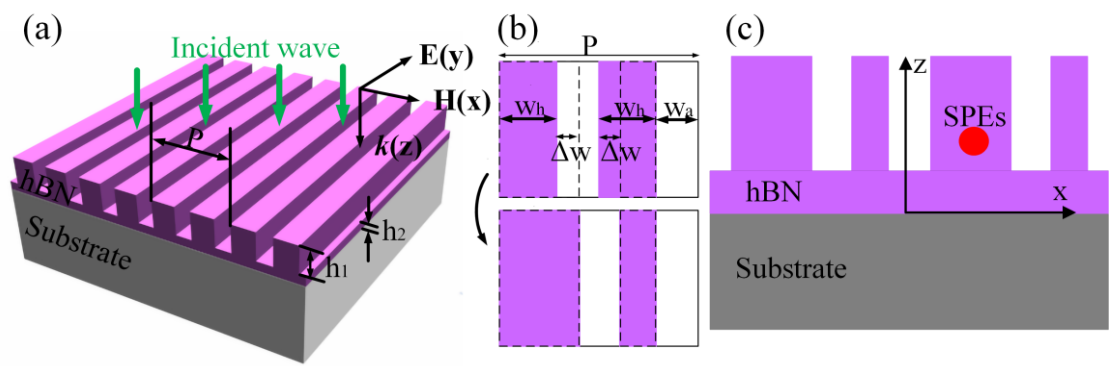

**Figure 1.** (a) Schematic of the unit cell of the monolithic hBN metasurface on a glass substrate. The inset shows the orientation and polarization of the incident wave. (b) Definition of the asymmetry parameter $\delta = \Delta w/w_h$ of the metasurface. (c) The sectional view of the monolithic hBN metasurface. The SPEs (marked as the red dot) is inside the hBN structure.

**3. Results and discussion**

**3.1. BIC and quasi-BIC modes in an hBN metasurface**

As shown in Fig. 1(a), we design a metasurface made of an hBN film and an hBN grating with broken in-plane inversion symmetry. The top layer is a two-part periodic grating with period P and thickness $h_1$. The unit cell of the hBN grating is composed of a pair of infinite bars, which have widths $w_h + \Delta w$ and $w_h - \Delta w$, respectively. The gap between any two neighboring bars has the same value $w_a$. The asymmetry parameter δ of the unit cell is defined as $\delta = \Delta w/w_h$, as illustrated in Fig. 1(b). The thickness of the lower hBN film is $h_2$. The hBN metasurface is placed on a silica (n = 1.46) substrate. Figure 1(c) is the sectional view of the proposed metasurface in the x-z plane. The quasi-BIC modes with large internal field enhancement (upto 350 times) can be supported with proper structural parameters (P = 400 nm, $h_1$ = 140 nm, $h_2$ = 46 nm, $w_a$ = 0.2 P, and $w_h$ = 0.3 P). Accordingly, due to the high Q/V ratio of the patterned structure we have desgined, the Purcell factor is considerably enlarged, which significantly enhances the spontaneous emission rate of the SPEs (marked as the red dot in Fig. 1(c)) inside the hBN structure. The SPEs in hBN can be generated by using thermal annealing or ion [12,18]. The effects of these methods in experiments on the structure fabrication and Q-values will be discussed later.

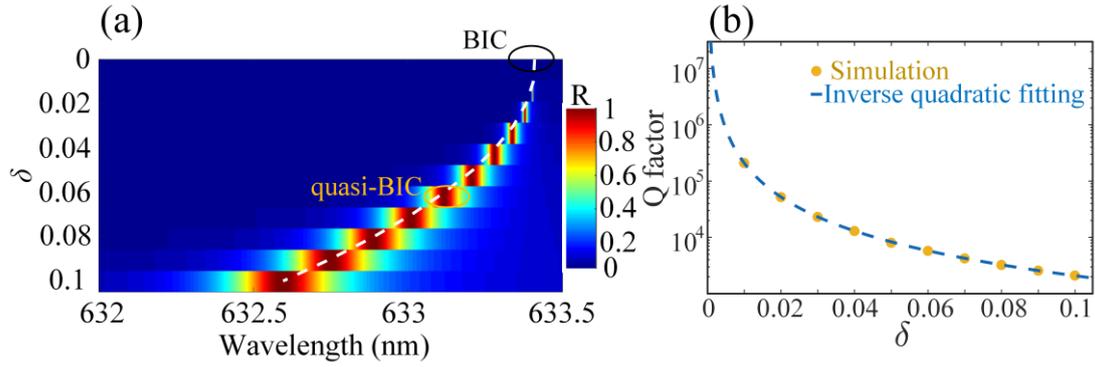

**Figure 2.** (a) Simulated reflectance spectra with respect to δ around the emission wavelength of 633 nm. The white dashed line in (a) illustrates the quasi-BIC dispersion. (b) Dependence of the Q factor on δ for quasi-BIC mode around 633 nm. The dashed line in (b) is inverse quadratic fitting lines.

The dependence of the calculated reflectance spectra on the wavelength around 633 nm and the asymmetry parameter δ are shown in Fig. 2(a). The white dashed line indicates the eigenmode dispersion of the metasurface. With δ decreases from 0.1 to near 0, it can be seen that the linewidth of the resonance decreases clearly and the resonance peak slightly shifts toward long wavelengths. When δ=0, the resonance linewidth vanishes completely. As shown by the eigenmode dashed lines, the symmetric hBN metasurface supports the symmetry-protected BIC mode at 633.41 nm, which have infinite Q value. When breaking the symmetry with δ > 0, unstable BIC modes transform into quasi-BIC modes with a finite Q factor. The quasi-BIC modes can be seen in the reflectance spectra with ultranarrow linewidth resonance. Furthermore, the corresponding Q factor of the quasi-BIC mode for different asymmetry parameter δ can be calculated by fitting the transmission spectra to Fano lineshape (see the Supplementary Material). When δ is 0.1, the Q factor of the quasi-BIC mode is 2089.1. Then the Q factor increases fast, when δ gradually approaches zero. For instance, when δ is 0.01, the Q factor reaches 208648.1, which is much larger than that of PCCs in refs. 9 and 12. As depicted in Fig. 2(b), the relationship between radiative Q factor and asymmetry parameter δ follows the inverse quadratic law, i.e., Q factor is proportional to $1/\delta^2$ [19]. For the experiments, one can vary δ from 0.01 to 0.1.

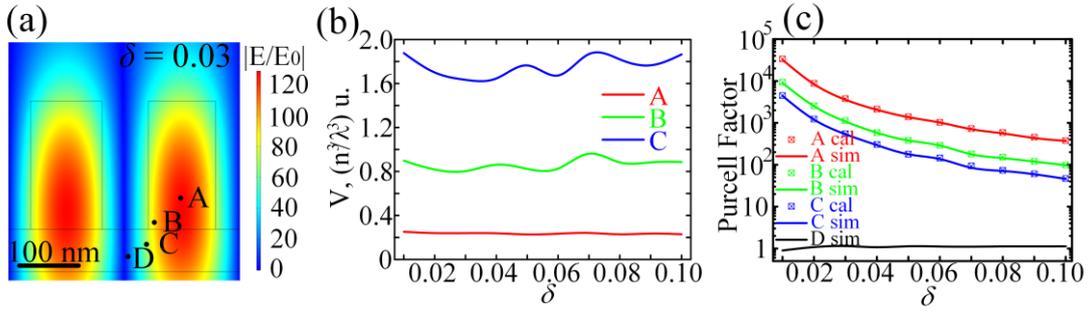

Figure 3. (a) The electric field profiles of the quasi-BIC mode for δ = 0.03 at the emission wavelength of 633 nm. The electric field enhancement factors are shown in the color bar. Points A, B, C and D represent the positions with different electric filed enhancements. (b) Normalized effective mode volume for different δ values at Points A, B and C. (c) The simulated and calculated maximal Purcell factors of an electric dipole emitter at different positions (Points A, B, C and D) plotted as a function of δ, respectively.

**3.2 Giant Purcell factor of single-photon emitters in the hBN metasurface with quasi-BIC modes**

The radiative decay rate of the quantum emitters can be engineered by the tailored local density of optical states (LDOS), which can also be optimized by a metasurface [20]. The electric field profiles of quasi-BIC modes around 633 nm in the xOz cut plane at the corresponding resonances under δ=0.03 is shown in Fig. 3(a), which also extends infinitely in the y-direction. The Figure S2 shows the electric field profiles of quasi-BIC modes under different δ (0.01-0.02 and 0.04-0.05) in the Supplementary Material. As can be seen from the figures, under different δ, the local electric field enhancement ($|E/E_0|$) is significantly different, although the electric field distribution appears similar. When δ=0.01, the maximum value of the electric field enhancement can exceed 350 times, which decreases significantly with the increase of δ (from 0.01 to 0.05). As shown in Fig. 3(a), four different points (A, B, C and D) are picked to represent different areas with different levels of electric filed enhancement. For δ=0.03, the electric field enhancements at Points A, B and C are around 100, 70 and 40 times, respectively. And the electric field at Point D has almost no enhancement. In the following simulations, the spatial positions of these four points are kept with different asymmetry parameter δ.

The Purcell factor of an electric dipole inside of the hBN metasurface is defined as [18]:

$$F = \frac{1}{8\pi}\left(\frac{\lambda}{n}\right)^3 \frac{Q}{V_{eff}} \quad (1)$$

where $V_{eff}$ is an effective mode volume and $\lambda$ is the resonant wavelength. The effective mode volume $V_{eff}$ can be calculated by [21-23]:

$$V_{eff}(\mathbf{r}_d) = \frac{\int_V \varepsilon(\mathbf{r})|\mathbf{E}(\mathbf{r})|^2 d\mathbf{r}}{\varepsilon(\mathbf{r}_d)[\mathbf{e}\cdot\mathbf{E}(\mathbf{r}_d)]^2} \quad (2)$$

where $\mathbf{r}_d$ and $\mathbf{e}$ are the position and polarization vector of the electric dipole, respectively. From Eq. (2), one can find that $V_{eff}$ is in inverse proportion to the local electric field intensity. Then $V_{eff}$ for different δ at three different positions (A, B and C) are calculated and the results are shown in Fig. 3(b). For different δ (from 0.1 to 0.01), the effective mode volume changes slightly, which together with the sharp increase in Q factor benefits to enhance the Purcell factor. Due to electric field enhancement at Point A is the highest among these three positions, $V_{eff}$ at Point A is the smallest when compared these three curves in Fig. 3(b).

The expected Purcell enhancement of single emitters due to the coupling to quasi-BIC modes are investigated by using 3D simulations of COMSOL Multiphysics. A quantum emitter inserted at point A, B, C, or D is simulated as a point electric dipole. The emission wavelength of the emitter dipole is assumed to be around 633 nm, which matches the experimentally studied hBN SPEs. The emitter dipole is oriented along the *y*-axis to match the polarization of the quasi-BIC mode. The Purcell factors ($\gamma_r / \gamma_r^0$) at the given positions were calculated as a ratio of spontaneous emission rate of a point dipole in the hBN metasurface ($\gamma_r$) and that in free space ($\gamma_r^0$). In experiments the physical size of the hBN metasurface is finite, and thus the hBN metasurface is modeled as N ✕ N arrays. The transmission spectrum and Q factor for N periods (N = 40, 70 and 100) of hBN bars are obtained (see the Supplementary Material). From Figure S3 in the Supplementary Material, one can find that when the number N reaches 100, the transmission and Q factor are very close to the case of infinite periods. Next, the hBN metasurface is modeled as 100 ✕ 100 arrays to simulate the Purcell factors. Perfectly matched layer (PML) boundary conditions surrounding the structure are employed. The maximum Purcell factors at points A, B, C and D with different δ (from 0.01 to 0.1; the corresponding wavelength varies for different δ; also see Fig. 4 below) are shown in Fig. 3(c). It can be found that the largest Purcell factor at Point A reaches up to 3.2 ✕ $10^4$ (δ=0.01), however, the Purcell factor decreases as the asymmetry parameter δ increases. Similar trends for the Purcell factor at Points B and C can also be observed from Fig. 3(c). For a specific δ, the Purcell factor at Point A is the largest because of its maximum enhancement of the electric field intensity among these four points. As a contrast, the Purcell factor for different δ at three different positions (A, B and C) are also calculated and shown in Fig. 3(c) by using Eq. (1). The calculated Purcell factor also show same characteristics as the simulated ones. It can be found that the simulated Purcell factor are slightly smaller than the calculated ones. This is because the Q factor of the 100 × 100 arrays is a little lower than the infinite metasurface. As almost no electric field enhancement at Point D, the Purcell factor is approximately equal to 1 and features no enhancement at Point D. Except for several individual parts in the lower hBN film, the electric field can be strongly enhanced in most parts of the present hBN structure. Therefore, this is beneficial to easy experimentally the precise positioning of the SPE in the hBN metasurface. For example, when the SPE is positioned into the top hBN grating layer, the expected Purcell enhancement is greater than 100 for δ=0.1.

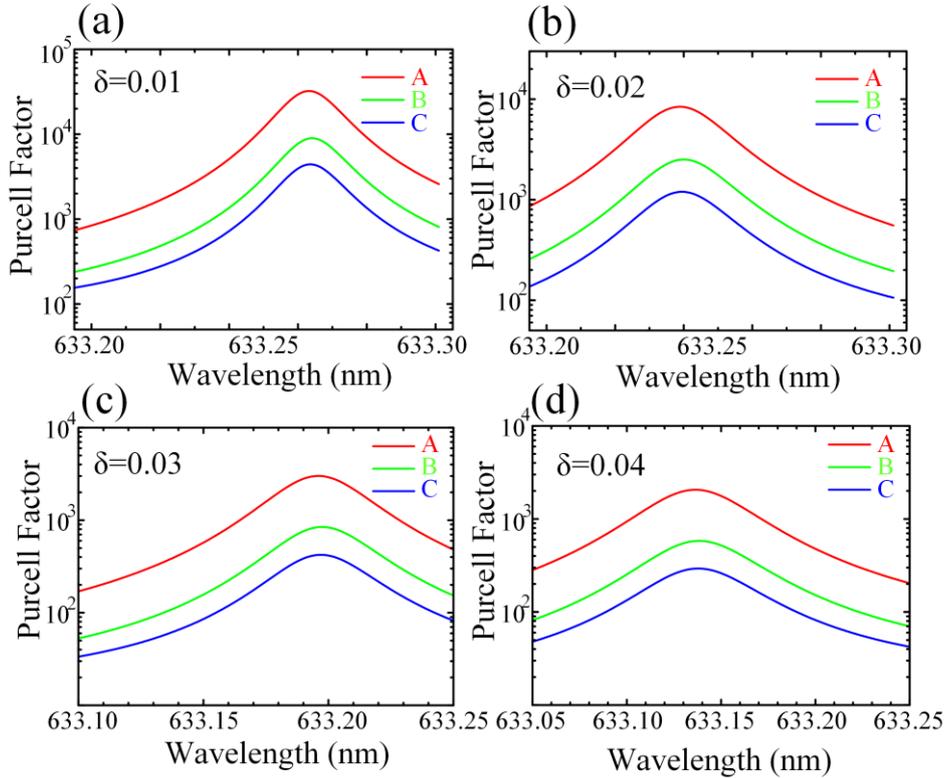

**Figure 4.** Purcell factors of an electric dipole emitter at different positions (Points A, B, C) as the wavelength varies for δ= (a) 0.01, (b) 0.02, (c) 0.03 and (d) 0.04.

The dependence of the Purcell factor on the emission wavelength of an electric dipole positioned at different points (Points A, B and C) and the asymmetry parameter δ (from 0.01 to 0.04) are shown in Fig. 4. For δ (from 0.05 to 0.08), the corresponding simulated Purcell factors are shown in Figure S6 in the Supplementary Material. It can be found that the full width at half maximum (FWHM) of the Purcell factor curve becomes narrower when δ is reduced from 0.08 to 0.01. For δ = 0.01, the FWHM is 0.018 nm, which means that its Q factor reaches 35181.1. Due to the coarser meshes and finite arrays used in simulations, this Q factor of Purcell factor is less than the Q factor of quasi-BIC mode (208648.15) (fitted by Eq. (S1); see Table S1 in Supplementary Material) when δ is 0.01. The Q factors of the Purcell factor curves are 21107.9, 14390.8, 9593.0 for δ =0.02, 0.03 and 0.04, respectively. Therefore, the emission of SPEs also has the property of ultra-narrow spectra. The wavelength of the Purcell factor peak exhibits blueshift when δ changes from 0.01 to 0.08, which accords with the blueshift of the quasi-BIC mode. It is worth noting that the wavelength of the Purcell factor curve has a small difference from the resonance wavelength of the quasi-BIC mode. This is because the mesh sizes for calculating the 3D Purcell factor of finite hBN arrays are a bit coarser (to save the calculation time) than those for calculating the 2D transmission spectra of infinite hBN metasurface.

For future experiments, two different methods can be used to precisely put the SPEs at positions where the electric field intensity of the quasi-BIC mode is maximal. One is to find pre-existing SPEs and post-fabricate top hBN gratings around them [24]. The other approach is to deterministically create emitters in desired locations [25]. The periodicity of the hBN structure will be broken to some degrees when ions are implanted for SPEs. Then the quality factor of the quasi-BIC modes and the electric field enhancement will become much weaker, and ultimately, the Purcell

Factor will also diminish. Hence, in order to create SPEs in hBN, an additional annealing step at 850 °C can be employed after the fabrication of the pattern structure [12], although the SPEs will be in random locations. As Purcell factor can be strongly enhanced in most parts of the hBN structure, thermal annealing is suitable to create SPEs in the present hBN metasurface. Annealing the hBN usually ends up creating a lot of defects in high concentration, which may also degrade the quality factor and field enhancement. In this work, we just considered only the Purcell Factor of single emitter, which could be fabricated by some local annealing or ions implantation. Due to the limitation of current nano-fabrications, the experimental quality factors of the hBN metasurface may be lowered by several-fold as compared to the numerical results. Then some fabrication errors, including the etched depth and width errors of the gratings, induced by nanofabricating processes are analyzed and discussed (see Figure S4 and S5 in the Supplementary Material for details). It is found that Q factor does not change noticeably by the following deviations: + or - 8 nm error for the etched depth and + or - 10 nm error for width of the grating. If the total number of the width imperfections is controlled under 10 out of 100 periods, Q factor will also change little.

In addition, the present work can also be extended into other relevant material platforms, such as quantum dots in III/V semiconductors and color centers in diamond. In future work, enhancing both the coupling efficiency and the emission of SPEs can be studied to make a useful source for quantum information processing. In general, the dipole orientations of emitters in 2D materials are random (see Ref. 26) and there will be no enhancements if the dipole orientations of the emitters are along x- or z-direction. One possible way to alleviate this problem is to design a patterned structure with polarization-independent quasi-BIC modes so that the emission of both x- and y-oriented emitters can be enhanced.

## 4. Conclusions

In conclusion, we have proposed and demonstrated a monolithic hBN metasurface by quasi-BIC mode with ultrahigh Q factor to enhance the signals emitted by SPEs in hBN. Quasi-BIC modes have been utilized to enhance the Purcell factor of the SPEs in hBN. Furthermore, the relationship between the radiative Q factor and Purcell factor of the SPEs has been carefully studied for various asymmetric parameters. Finally, an ultrahigh Purcell factor of $3.3 \times 10^4$ has been achieved. Because the electric field can be strongly enhanced in most areas of the proposed hBN structure, it can ease the experimental conditions to precisely position the SPEs in the hBN metasurface. Our proposed monolithic hBN metasurface with ultrahigh Q values and Purcell factor may find potential applications from onchip single-photon sources and quantum circuits, to sensing biochips, high-harmonic generations and fluorescence enhancement of e.g. upconversion nanoparticles.

**Abbreviations**

hBN: hexagonal boron nitride; SPEs: single photon emitters; PCCs: photonic crystal cavities; BICs: bound states in the continuum; LDOS: local density of optical states; PML: Perfectly matched layer; FWHM: full width at half maximum;

**Declaration**

**Acknowledgements**

Not applicable.

**Authors' contributions**

SC and SH conceived the idea. SC performed the simulations, analyzed the data, created the figures and finished the original version of the manuscript. All authors participated in the discussion, revised the manuscript and approved the final manuscript. SH supervised the whole work. All authors approved the final manuscript.

**Competing interests**

The authors declare no competing financial interest.

**Funding**

This work was partially supported by the National Natural Science Foundation of China (No. 91833303, 11621101, 61774131, 61875174), the National Key Research and Development Program of China (No. 2017YFA0205700) and Zhejiang Provincial Natural Science Foundation of China (LY17F010006)

**Availability of data and materials**

The datasets generated and/or analyzed during the current study are available from the corresponding author on reasonable request.

# References

1. M. Autore, P. Li, I. Dolado, F. J Alfaro-Mozaz1, R. Esteban, A. Atxabal, F. Casanova, L. E Hueso, P. Alonso-González, J. Aizpurua, A. Y Nikitin, S. Vélez, R. Hillenbrand, Light Sci. Appl. **7**, 17172 (2018)

2. T.-T. D. Tran, V. W. Bray, M. J. Ford, M. Toth, I. Aharonovich, Nat. Nanotechnol. **11**, 37–41 (2016)

3. A. Ambrosio, M. Tamagnone, K. Chaudhary, L. A. Jauregui, P. Kim, W. L. Wilson, F. Capasso, Light Sci. Appl. **7**, 27 (2018)

4. T. T. Tran, C. ElBadawi, D. Totonjian, G. Gross, H. Moon, D. Englund, M. J. Ford, I. Aharonovich, M. Toth, ACS Nano **10**, 7331 (2016)

5. M. Nguyen, S. Kim, T. T. Tran, Z.-Q. Xu, M. Kianinia, M. Toth, I. Aharonovich, Nanoscale **10**, 2267–2274 (2018)

6. T. Trong Tran, D. Wang, Z.-Q. Xu, A. Yang, M. Toth, T. W. Odom, I. Aharonovich, Nano Lett. **17**, 2634–2639 (2017)

7. T. Vogl, R. Lecamwasam, B. C. Buchler, Y. Lu, P. K. Lam, ACS Photonics **6**, 1955−1962 (2019)

8. S. Kim, N. M. H. Duong, M. Nguyen, T. Lu, M. Kianinia, N. Mendelson, A. Solntsev, C. Bradac, D. R. Englund, I. Aharonovich, Adv. Optical Mater. 1901132 (2019)

9. J. E. Fröch, S. Kim, N. Mendelson, M. Kianinia, M. Toth, I. Aharonovich, ACS Nano **14**, 7085-7091 (2020)

10. G. Cassabois, P. Valvin, B. Gil, Nat. Photonics **10**, 262 (2016)

11. Z. Liu, Y. Gong, W. Zhou, L. Ma, J. Yu, J. C. Idrobo, J. Jung, A. H. MacDonald, R. Vajtai, J. Lou, P. M. Ajayan, Nat. Commun. **4**, 2541 (2013)

12. S. Kim, J. E. Fröch, J. Christian, M. Straw, J. Bishop, D. Totonjian, K. Watanabe, T. Taniguchi, M. Toth, I. Aharonovich, Nat. Commun. **9**, 2623 (2018)

13. J. E. Fröch, Y. Hwang, S. Kim, I. Aharonovich, M. Toth, Adv. Opt. Mater. **7**, 1801344 (2018)

14. C. W. Hsu, B. Zhen, A. D. Stone, J. D. Joannopoulos, M. Soljačić, Nat. Rev. Mater. **1**, 16048 (2016)

15. S. Cao, H. Dong, J. He, E. Forsberg, Y. Jin, S. He, J. Phys. Chem. Lett. **11**, 12, 4631-4638 (2020)

16. S. Kolodny, I. Iorsh, Opt. Lett. **45**, 181-183 (2020)

17. Z. Liu, Y. Xu, Y. Lin, J. Xiang, T. Feng, Q. Cao, J. Li, S. Lan, J. Liu, Phys. Rev. Lett. **123**, 253901 (2019)

18. S. Choi, T. T. Tran, C. Elbadawi, C. Lobo, X. Wang, S. Juodkazis, G. Seniutinas, M. Toth, I. Aharonovich, ACS Appl. Mater. Interfaces **8**, 29642-29648 (2016)

19. K. Koshelev, S. Lepeshov, M. Liu, A. Bogdanov, Y. Kivshar, Phys. Rev. Lett. **121**, 193903 (2018)

20. C. Gong, W. Liu, N. He, H. Dong, Y. Jin, S. He, Nanoscale **11**, 1856-1862 (2019)

21. E. Muljarov, W. Langbein, Phys. Rev. B **94**, 235438 (2016)

22. R. Coccioli, M. Boroditsky, K. W. Kim, Y. Rahmat-Samii, E. Yablonovitch, IEE Proc.-Optoelectron. **145**, 391 (1998)

23. Heiko Groß, J. M. Hamm, T. Tufarelli, O. Hess, B. Hecht, Sci. Adv. **4**, eaar4906 (2018)

24. M. Kianinia, C. Bradac, B. Sontheimer, F. Wang, T. T. Tran, M. Nguyen, S. Kim, Z.-Q. Xu, D. Jin, A. W. Schell, C. J. Lobo, I. Aharonovich, M. Toth, Nat. Commun. **9**, 874 (2018)

25. Z.-Q. Xu, C. Elbadawi, T. T. Tran, M. Kianinia, X. Li, D. Liu, T. B. Hoffman, M. Nguyen, S. Kim, J. H. Edgar, X. Wu, L. Song, S. Ali, M. Ford, M. Toth, I.

Aharonovich, Nanoscale **10**, 7957-7965 (2018)

26. F. Peyskens, C. Chakraborty, M. Muneeb, D. Van Thourhout, D. Englund, Nat. Commun. **10**, 4435 (2019)